\documentclass[twocolumn]{IEEEtran}
\usepackage{cite}
\usepackage{graphicx}
\usepackage[colorlinks=true,linkcolor=red,urlcolor=blue,citecolor=red]{hyperref}
\usepackage{upgreek}
\usepackage[colorinlistoftodos]{todonotes}

\begin{document}

\title{\bf Magnetic Uncertainties for Compact Kibble Balances: An Investigation}

\author{Shisong Li$\dagger$,~\IEEEmembership{Senior Member,~IEEE,}
        Stephan Schlamminger$\ddagger$,~\IEEEmembership{Senior Member,~IEEE} 
\thanks{S. Li is with the Department of Electrical Engineering, Tsinghua University, China. S. Schlamminger is with the National Institute of Standards and Technology (NIST), United States.}%
\thanks{$\dagger$ shisong.li@outlook.com;$\ddagger$stephan.schlamminger@nist.gov} 
\thanks{Submitted to IEEE Trans. Instrum. Meas.}}


\maketitle

\begin{abstract}
	The Kibble balance has become one of the major instruments for realizing the mass unit, the kilogram, in the revised international system of units (SI). Researchers at about a dozen national metrology institutes are actively working with Kibble balances that are capable of weighing masses with nominal values from 10\,g to 1\,kg. In the future, the design of smaller Kibble balances will play a more significant role. Smaller Kibble balances require smaller magnet systems, and here we investigate the scaling of systematic uncertainties with the size of the magnet system.  We describe the size dependence of three magnetic effects: the coil-inductance effect, the yoke nonlinear effect, and the thermal effect. The analysis shows that the relative systematic effects become increasingly larger with smaller sizes. For small magnets the thermal effects become dominant and, hence, a good thermal design is imperative. 
\end{abstract}

\begin{IEEEkeywords}
Kibble balance, magnetic error, coil inductance, magnetic field measurement, thermal effect.
\end{IEEEkeywords}
\IEEEpeerreviewmaketitle

\section{Introduction}
\IEEEPARstart{T}{he} Kibble balance \cite{Kibble1976} compares the mechanical watt to the electric watt in terms of quantum measurements \cite{haddad2016bridging}, and can link the unit of mass, i.e. kilogram, to the Planck constant $h$ with relative uncertainties of a few parts in $10^8$. In the revised international system of units (SI), the Kibble balance is one of the major instruments for realizing the mass at the kilogram level and {below} \cite{NIST,NRC,METAS,LNE,BIPM,NIM,KRISS,MSL,UME,NPL3}. Details of the Kibble balance can be found in recent review papers in the field, for example, \cite{Stephan16}.

The measurements with a Kibble balance are conducted in two phases, the weighing phase and the velocity phase. The results of both are connected via a physical quantity, the so-called 'geometrical factor' $Bl$ ($B$ denotes the magnetic flux density at the coil position and $l$ the wire length of the coil): In the weighing phase, a current $I$ is injected into the coil placed in a magnetic field $B$. From the magnetic force and  the weight of a test mass $mg$ the geometric factor is determined as
\begin{equation}
    (Bl)_w=\frac{mg}{I},
\end{equation}
where $g$ is the local gravitational acceleration at the mass position. In the velocity phase, the $Bl$ is self-calibrated by moving the coil in the same magnetic field, obtaining
\begin{equation}
    (Bl)_v=\frac{U}{v},
\end{equation}
where $U$ is the induced voltage on the coil terminals and $v$ is the vertical velocity of the coil. In the {ideal case},  $(Bl)_w$ and $(Bl)_v$ are the same physical quantity and can be eliminated. Hence, the mass $m$ can be written as
\begin{equation}
    m=\frac{UI}{gv}.
    \label{eq3}
\end{equation}
In practice, however, the $Bl$ measurement can be affected by many factors, such as the magnetic field quality and stability, effectively leading to a $(Bl)_w$ that differs slightly from $(Bl)_v$. In this case, a measurement bias is introduced {in} the mass determination. The relative bias is given by
\begin{equation}
    \eta=\frac{\Delta m}{m}=\frac{(Bl)_w}{(Bl)_v}-1.
    \label{eq4}
\end{equation}
In a Kibble-balance measurement, $\eta$ {should} be smaller than the desired measurement uncertainty, typically a few parts in $10^8$. Although, the term $Bl$ is absent in (\ref{eq3}) and it seems that the mass measurement is independent of the magnet-coil system, it is not the case because in order to obtain (\ref{eq3}), the equivalence of $(Bl)_w$ and $(Bl)_v$ was already used. As presented in (\ref{eq4}), any imperfect $Bl$ cancellation will result in a measurement bias for the mass determination. 

During the last few decades, many different magnet systems were developed, e.g. \cite{nist3,LNEmag,BIPMmag2017,zhang2015coils,NISTmag,MSL}. Nowadays, the yoke-based permanent magnet system has been widely accepted and is a dominant component in most Kibble balances. {Among} many different magnet realizations, the BIPM-type design is used predominantly\cite{BIPMmag2006,NISTmag,METAS,you2016designing,KRISS,UME}. The advantages of using a   permanent magnet system include a high efficiency for generating a strong and uniform magnetic field, good robustness for long-term operation, and low maintenance cost. Systems with a closed yoke, such as the BIPM design, shield the coil from ambient magnetic fields. Only two disadvantages of  permanent magnet systems are noteworthy: First, the magnetically soft material of the yoke interacts with the coil current during the weighing measurement, and its nonlinear property can introduce systematic effects \cite{linonlinear,linonlinear2,hysteresis}; Second, the rare-earth materials that are used for the permanent magnet, such as Samarium Cobalt (SmCo), have sizeable temperature coefficients (about $-3\times10^{-4}${/K} for SmCo). Thus, thermal effects, especially those caused by ohmic heating during the weighing mode must be carefully considered.

The conventional permanent magnet systems used in Kibble balance experiments are usually big and heavy. For example, the NIST-4 magnet weights about 850\,kg\cite{NISTmag}, and the NIM-2 magnet system 500\,kg\cite{NIM}. One of the most exciting trends for future Kibble balances is toward smaller, more compact devices.
In the recent past, several national metrology institutes (NMIs) have launched tabletop Kibble balance projects that aim to realize masses from grams to kilograms~\cite{NPL3,chao2019design,chao2020performance,rothleitner2020planck,marangoni2019magnet}. 
One major thrust of the work is designing a smaller magnet system, ideally, systems that weigh only a few tens of kilograms or less.

Our previous studies, however, have indicated that the biases and uncertainties associated with the magnet system, have a strong negative dependence on the size of the magnet and coil, especially the volume of the air gap \cite{NISTmag,BIPMmag2017,li17,li18,linonlinear,linonlinear2,hysteresis}. 
Accordingly, we are interested in finding the size limit for the magnet system. How small can the permanent magnet be for a given desired relative uncertainty and mass value? {State of the art} values are $2\times10^{-8}$ for the relative uncertainty and 1\,kg for the mass of the test mass.
In this article, we investigate the most important factors that limit the down-scaling of the magnet systems, and we will discuss the mechanism of how these factors affect the measurement bias. 
Three major magnetic effects, as a function of the magnet volume, are discussed respectively in sections \ref{sec03}, \ref{sec04}, and \ref{sec05}. 

\section{General Considerations}
\label{sec02}

\begin{figure}
    \centering
    \includegraphics[width=0.45\textwidth]{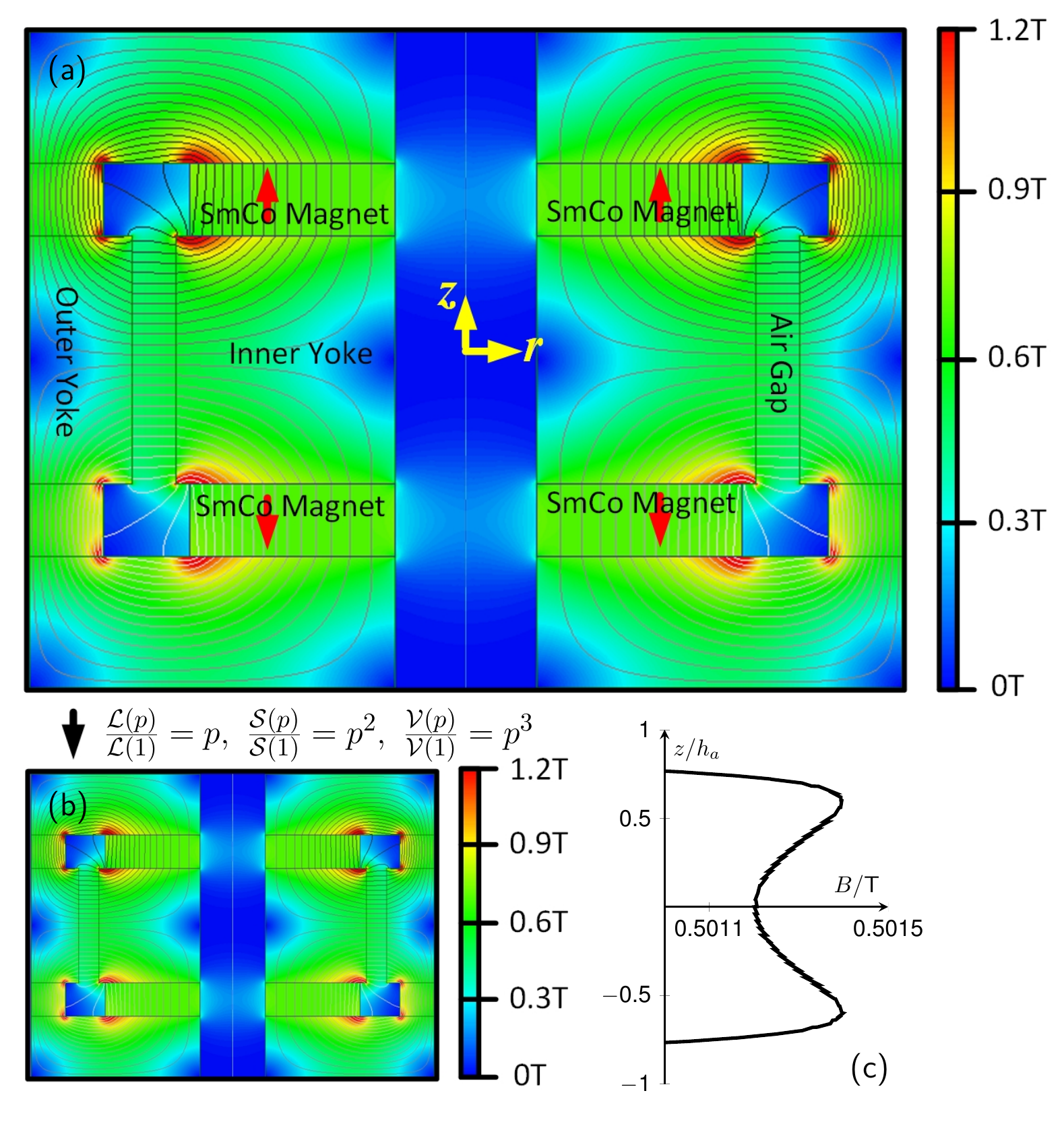}
    \caption{(a) shows the magnetic flux distribution in a typical BIPM-type Kibble balance magnet, NIST-4 \cite{NISTmag}. The magnetic flux of two SmCo permanent magnets is guided through a radial air gap. The magnetic field distribution along the vertical direction in the air gap has a good uniformity, which is a requirement  for the measurements in both phases. (b) shows the magnetic flux density distribution in a scaled magnet. (c) presents the $B(z)$ curve with a relative displacement axis, $z/h_a$, where $h_a$ is half of the air gap height.}
    \label{fig01}
\end{figure}

The main goal of this article is to investigate how the magnet-related measurement error scales with the magnet size or volume. Hence, the introduction of a scaling factor is of the essence. We use $p$ as a linear scale factor, any given side of the magnet scales with $p$. In this paper, the NIST-4 magnet system \cite{NISTmag} is used as a reference with $p=1$, and its total mass $M(p=1)=850$\,kg. When the volume of the system is reduced, the scaling factor in length, surface (or sectional area), and volume (or mass), respectively denoted by $\mathcal{L}(p)$, $\mathcal{S}(p)$ and $\mathcal{V}(p)$, are 
\begin{equation}
    \frac{\mathcal{L}(p)}{\mathcal{L}(1)}=p,~~\frac{\mathcal{S}(p)}{\mathcal{S}(1)}=p^2,~~\frac{\mathcal{V}(p)}{\mathcal{V}(1)}=p^3.
\end{equation}
For example, the magnet masses obtained for $p=1/3$ and $p=1/4$, $M(1/3)\approx31$\,kg and $M(1/4)\approx13$\,kg, are {quite} suitable for integration  in tabletop Kibble balances.   

As shown in Fig. \ref{fig01}, without changing the internal design, 
{ the geometrical scale factor $p$ alters the magnetic flux source and the flux path length with the same factor (for the line integral, it is $p$); }
hence the magnetic field in the air gap remains unchanged, or, in the notation introduced above,
\begin{equation}
    \frac{B(p)}{B(1)}=1.
\end{equation}
Experimental proof of this fact is the magnet system constructed at the national metrology institute, Turkey, named TÜBİTAK-Ulusal Metroloji Enstitüsü (UME). The UME Kibble balance built a magnet using the  NIST-4 designs {and} scaling it by  $p=1/3$~\cite{UME}. As expected, the resulting magnet has a similar magnetic flux density in the gaps as the original design.

The desired value for the geometric factor is a trade-off in the performance in weighing and velocity mode. A small $Bl$ value yields a low induced voltage $U$ and, thereby a larger relative uncertainty of the  $U$ measurement. On the other hand, a large $Bl$ value requires a small current for the weighing, and the relative measurement uncertainty for the current will increase. As detailed in \cite{schlamminger2013design}, an optimal $Bl$ value exists and it is  
\begin{equation}
    (Bl)_{op}=\sqrt{\frac{mgR}{v}},
    \label{eq:opBl}
\end{equation}
where $R$ is the resistance used in the electric circuit to  measure currents in the weighing phase. In this article, we assume that the magnet system is operated at $(Bl)_{op}$ independent of $p$. Since $B(p)=B(1)$ and $m$, $R$, and $v$ are all independent of $p$, so must be the length of the wire, i.e., $l(p)=l(1)$. In order for this assumption to hold since the coil circumference is proportional to $p$, $2\pi r(p)=p2\pi r(1)$, the number of windings must scale inversely to $p$, $N(p)=N(1)/p$. Another consequence of the assumption of constant $Bl$ as a function of $p$ is that the current required to measure the same mass is also independent of $p$. In summary,
\begin{equation}
    \frac{l(p)}{l(1)}=1,~~\frac{I(p)}{I(1)}=1~~\frac{N(p)}{N(1)}=\frac{1}{p}.
\end{equation}

For the next three sections assume $m$ to be constant at $m=1\,$kg. In the section after that, we also scale the mass of the test mass.

\section{Coil Inductance Effect}
\label{sec03}
\begin{figure}
    \centering
    \includegraphics[width=0.45\textwidth]{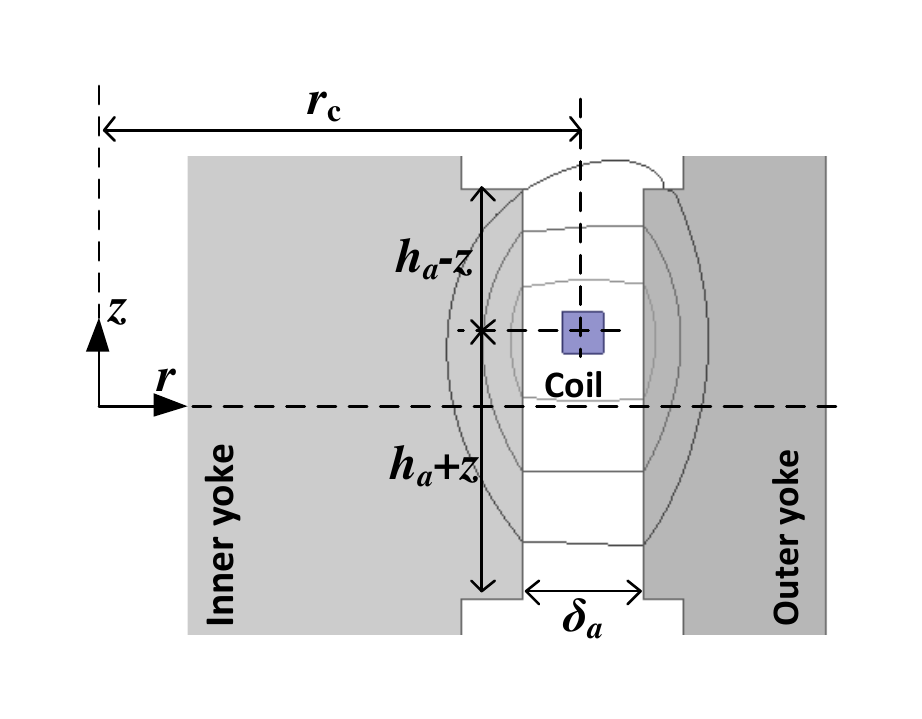}
    \caption{The dimension of the air gap. The circles present the magnetic flux lines when the coil locates at $z$. }
    \label{fig0x}
\end{figure}

For the conventional measurement scheme, a current in the coil is only present  in the weighing measurement. A coil current $I$ produces an additional flux loop returning through the air gap, introducing correction terms to the main field and hence the $(Bl)_w$ measurement. The mathematical model of the effect of the current on $(Bl)$ was originally postulated by Ian Robinson \cite{Robinson2007An}. He used
\begin{equation}
    \frac{(Bl)_w}{(Bl)_v}\approx1+\alpha I+\beta I^2,
    \label{eq:ian}
\end{equation}
where $\alpha$ and $\beta$ denote the linear and quadratic  coefficients, respectively. The inductance effect is a consequence of the linear part of the coil current effect ($\alpha I$). Usually, the weighing measurement is a two-part measurement, one with and one without the mass on the balance pan, labeled mass-on and mass-off, respectively. The balance tare is set such that the currents are symmetrical, $I_\mathrm{on}=-I_\mathrm{off}=I$. In this case, the linear term cancels by combining the two measurement results. However, experimental studies found that $\alpha$ has considerable dependence on the coil position $z$ \cite{li17} because a (large) part of it is generated by the gradient of the coil inductance $L$ along the vertical direction, i.e.
\begin{equation}
    \Delta(Bl)=\frac{\Delta F}{I}=\frac{I}{2}L'\rightarrow \alpha=\frac{L'}{2(Bl)},
\end{equation}
where $L'=\partial L/\partial z$ {and $\Delta F$ is the difference in force produced by the magnet system for the coil position $z$ and 0. As shown in Fig.\ref{fig0x}}, the coil flux threads the air gap twice, and the coil inductance can be written as 
\begin{equation}
    L=\frac{\mu_02\pi r_aN^2}{2\delta_ah_a}(h_a^2-z^2),
    \label{eq:L}
\end{equation}
and hence $\alpha$ is given by
\begin{equation}
    \alpha=-\frac{\mu_02\pi r_aN^2}{2\delta_ah_a(Bl)}z,
    \label{eq:alpha}
\end{equation}
where $N$ is the number of wire turns in the coil, $r_a$ the mean radius of the air gap, $\delta_a$ the air gap width, and $h_a$ the equivalent half-height of the air gap ($h_a=\gamma h_\mathrm{geo}$ \cite{diamagnetic2020} where $h_\mathrm{geo}$ is the geometrical height and $\gamma$ is a constant with $\gamma>1$). The major mechanism leading to a systematic bias during mass-on and mass-off measurements is that the coil suspensions expands or contracts with the load change. Hence, if the balance beam 
is {servoed} to the same position, the coil position will change between mass-on and mass-off. The difference $z_\mathrm{on}-z_\mathrm{off}$ can range from a few micrometer to more than ten micrometer. The weighing position change yields to a bias term given by
\begin{equation}
\varepsilon_1=\frac{\alpha(z_\mathrm{on})I_\mathrm{on}+\alpha(z_\mathrm{off})I_\mathrm{off}}{2}.
\label{eq:e1complicated}
\end{equation}
Using $I_\mathrm{on}=-I_\mathrm{off}=I$, (\ref{eq:alpha}) and $l=2\pi r_aN$, (\ref{eq:e1complicated}) can be simplified to
\begin{eqnarray}
    \varepsilon_1&=&-\frac{\mu_0NI}{4\delta_ah_aB}(z_\mathrm{on}-z_\mathrm{off})\nonumber\\
    &=&\frac{L''(z_\mathrm{on}-z_\mathrm{off})I}{4(Bl)},
    \label{eq:e1}
\end{eqnarray}
where $L''=\partial^2 L/\partial z^2$.
In (\ref{eq:e1}), $Bl$, $B$ and $I$ are independent on $p$,  and hence the volume dependence of the inductance error $\varepsilon_1$ is determined by $N$, $\delta_a$, and $h_a$. Since $N(p)=N(1)\,p^{-1}$, $\delta_a(p)=\delta_a(1)\,p$, and $h_a(p)=h_a(1)\,p$, we find
\begin{equation}
    \frac{\varepsilon_1(p)}{\varepsilon_1(1)}=p^{-3}.
    \label{eq:e1r}
\end{equation}

Eq.~(\ref{eq:e1r}) is obtained assuming that the change in coil position between the mass-on and mass-off measurement remains the same. That would be one extreme assumption. The other extreme would be, that $\Delta z (p) = \Delta z (1) p^{-1}$. In this case, $\varepsilon_1(p)=\varepsilon_1(1)p^{-4}$. The reality is probably in the middle of these two cases.

The scaling of the inductance error $\varepsilon_1$ is given in (\ref{eq:e1r}), but to compare its size to that of other magnet errors, $\varepsilon_1(1)$ needs to be determined.
We start with (\ref{eq:e1}) and use values of $L''$, $I$ and $Bl$ to estimate $\varepsilon_1(1)$ as a function of $z_\mathrm{on}-z_\mathrm{off}$. The current $I=6.9$\,mA and $Bl\approx710$\,Tm are known. The difficulty lies in finding $L''$.  Table~\ref{tab:L} lists  $L''$ determined by several different methods. The FEA (finite element analysis) calculates directly the inductance force with $\Delta F=\frac{I^2}{2}L'$, and hence $L''=\frac{2}{I^2}\frac{\partial \Delta F}{\partial z}$. The value of $L''$ can also be deduced form experimental measurements. Reference \cite{NIST} gave a determination $L''=(-656\pm332)$\,H/m$^2$ by measuring the Planck constant $h$ with different masses. However, the measurement accuracy is limited to about 50\%. Another experimental measurement by directly measuring $L$ was carried out in \cite{NISTmag}, It yielded $L''=-346$\,H/m$^{2}$ with a 928-turn test coil. The new measurement coil of NIST-4 experiment has 945 turns and hence $L''=-346\times(945^2/928^2)=-358.8$\,H/m$^{2}$. Note that this result has a much smaller uncertainty than the value determined by $h$ measurement.

\begin{table}[tp!]
    \centering
    \caption{Results of the $L''$ determination.}
    \begin{tabular}{clc}
    \hline
    method & $L''$/(Hm$^{-2}$) & uncertainty/(Hm$^{-2}$) \rule[-4pt]{0pt}{12pt}\\
    \hline
    FEA	    &$-335.7$	&not available \\
    Exp. ($h$ meas.)	&$-656$	&332\\
    Exp. ($L$ meas.)	&$-358.8$	&not available  \\
    \hline
    \end{tabular}
    \label{tab:L}
\end{table}

\begin{figure}
    \centering
    \includegraphics[width=0.45\textwidth]{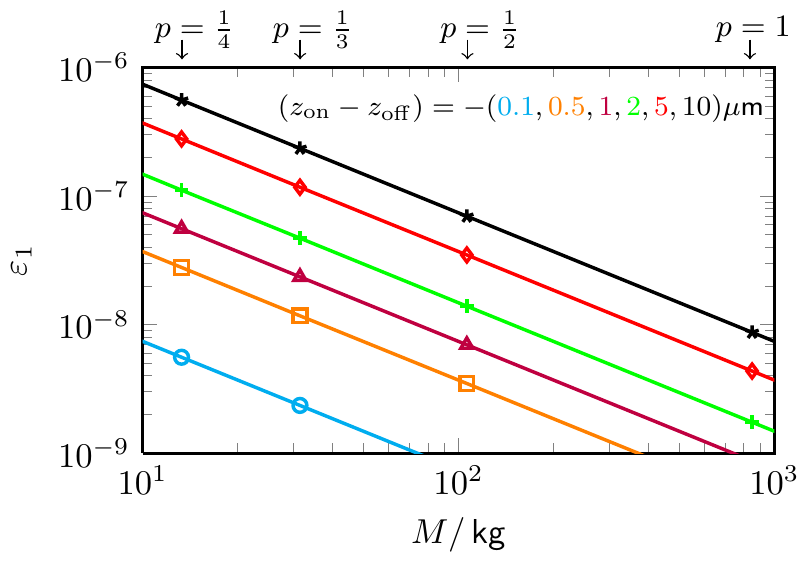}
    \caption{The relative inductance effect $\varepsilon_1$ with different values of $z_\mathrm{on}-z_\mathrm{off}$ as a function of magnet mass. The $p$ values are marked on the upper horizontal axis.}
    \label{fig:e1}
\end{figure}

We use the latter $L$ measurement result to evaluate $\varepsilon_1(1)$ and obtain $\approx-9\times10^{-10}$ for $z_\mathrm{on}-z_\mathrm{off}= 1\,\upmu$m. With this starting {point} and this calling given in (\ref{eq:e1r}), the coil-inductance error for different values $z_\mathrm{on}-z_\mathrm{off}$ is calculated. The result is shown in Fig.~\ref{fig:e1}. As mentioned before, a change in $z$ is inevitable because during weighing the balance beam is controlled to a null position, but the finite stiffness of the coil stirrup causes a change in the coil position as the produced force changes from $mg/2$ in the mass-off to $-mg/2$ in the mass-on measurement. In the NIST-4 system, the spring constant for the suspension was measured as $\kappa=0.7$\,N/$\upmu$m, and hence 1\,kg mass gravity, $mg$, gives $z_\mathrm{on}-z_\mathrm{off}=-14\,\upmu$m or a correction of $1.26\times10^{-8}$ for the mass measurement result. Usually, $\varepsilon_1$ is significant and needs to be carefully considered. The effect becomes larger as the magnet gets smaller. For example for $p=1/3$, the inductance correction is $\approx 10^{-7}$ and a correction must be applied to the result. 

\section{Magnetic Nonlinear Effect}
\label{sec04}
The nonlinear magnetic effect refers to the quadratic term in (\ref{eq:ian}), i.e. $\beta I^2$. Unlike the linear term, the quadratic term can not be eliminated by combining the results obtained with the mass-on and mass-off  measurements. Because this term is proportional to $I^2$, changing the sign of the current will produce the same change of the magnetic field, $\Delta H_y^2$. At the weighing position, which is ideally at $z=0$ where the magnetic field profile is flat, the coil flux returns in two halves of the air gap. Ignoring the magneto-motive force drop in the yoke, the Ampere's law yields
\begin{equation}
2\Delta H_a \delta_a=NI~~\rightarrow \Delta H_y=\frac{H_a}{\mu_r}=\frac{NI}{2\mu_r\delta_a}.\label{eq:MNL}
\end{equation}
Since $N\propto p^{-1}$ and $\delta_a\propto p$, and hence $\Delta H_y\propto p^{-2}$. Then the nonlinear magnetic error, which scales proportionally to $\Delta H_y^2$ is
\begin{equation}
\frac{\varepsilon_2(p)}{\varepsilon_2(1)}=\frac{\Delta H_y^2(p)}{\Delta H_y^2(1)}=\frac{1}{p^4}.
\end{equation}

In general, that additional magnetic field can produce a systematic effect in two ways. First, a part of the coil flux enters the yoke and moves its working point along the $B-H$ curve of the yoke. A detailed description of this effect has been given in \cite{linonlinear, linonlinear2}. Both components, the field that is parallel and perpendicular to the main magnetic flux were analyzed. In the end, both studies yielded the relationship $\beta I^2\propto (\Delta H_{y})^2$. The theoretical investigation predicted that the nonlinear term $\beta I^2$ is below $1\times10^{-9}$ for the NIST-4 system.   

\begin{figure}
    \centering
    \includegraphics[width=0.5\textwidth]{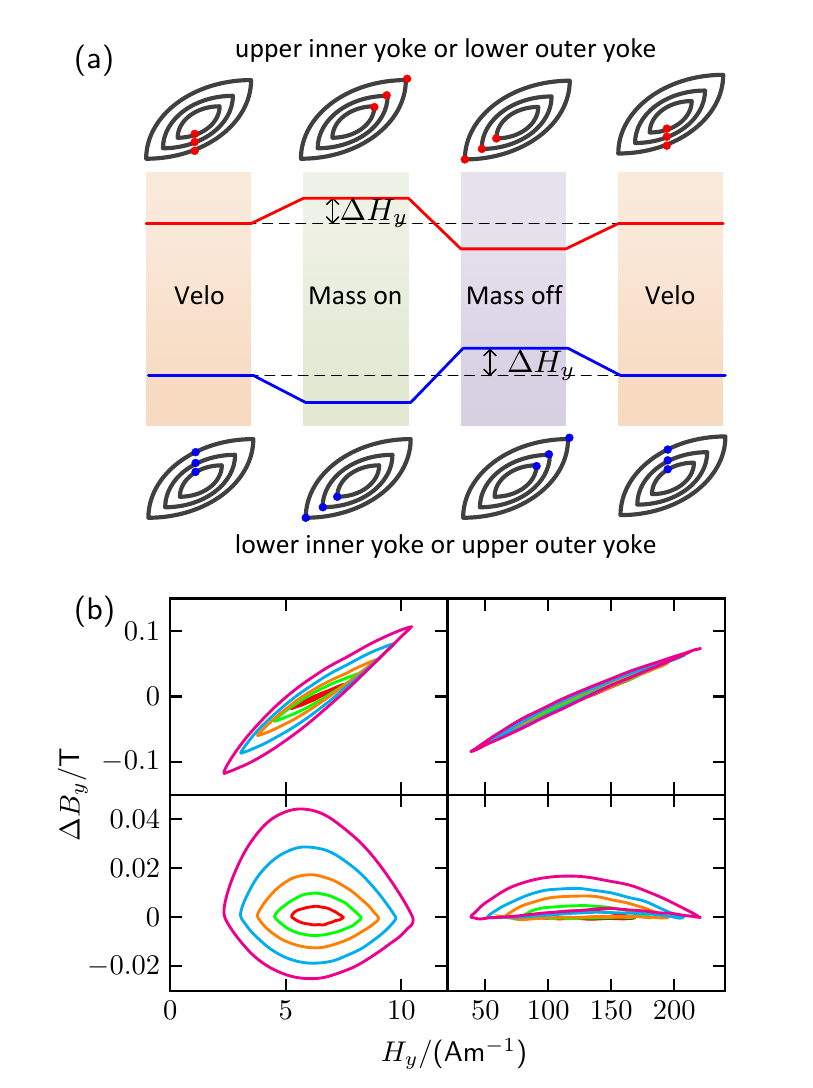}
    \caption{(a) shows the working points of different yoke parts on the hysteresis curve for a complete cycle of Kibble balance measurement. The red and blue curves present the $H$ field change of the yoke caused by the weighing current.  The current ramps before each weighing measurement. The $BH$ loops presented are minor loops and the blue and red dots show the working points. (b) shows an experimental determination of the minor loop of soft yoke materials at the BIPM Kibble balance group. The graphs in the left column were measurements taken on material that was heat treated. The data in the right column were taken with samples that were not heat treated. The graphs in the first row show the raw data. The graphs in the second row were obtained by subtracting a linear slope from the data in the first row.}
    \label{fig:hys}
\end{figure}

The second mechanism for producing a nonlinear magnetic error is related to the hysteresis of the yoke.  The working points of each quarter of the yoke during three phases of the Kibble balance measurement are shown  in Fig. \ref{fig:hys} (a).
The equivalent magnetic flux density for the velocity measurement equals the average of the upper and lower $B_y$ where $\Delta H_y=0$. In the weighing measurement, the $B$ field measured is the average of the upper and lower endpoints ($\Delta H_y$ and -$\Delta H_y$). In a perfect symmetric scenario, no bias occurs. In reality, however, that symmetry is broken by the non-reversal change in the magnetization given by the minor hysteresis loops. Fig. \ref{fig:hys} (b) shows an experimental measurement of minor hysteresis loops of a solid yoke material that was biased with a $B$ field \cite{hysteresis}, the $BH$ loop is no longer symmetrical. Hence, the average $B$ field in the weighing and  that in velocity measurements are no longer the same. 

\begin{figure}
    \centering
    \includegraphics[width=0.45\textwidth]{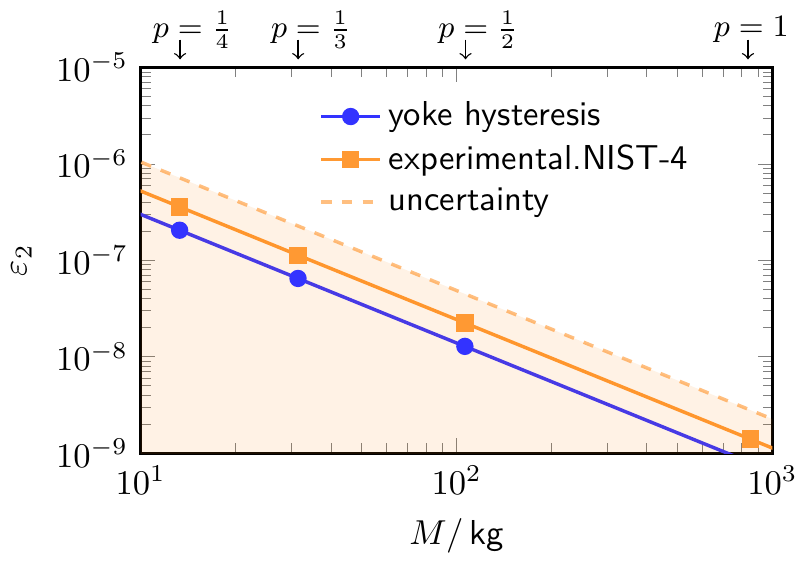}
    \caption{The nonlinear magnetic error as a function of magnet mass (lower horizontal axis) and $p$ (upper horizontal axis).}
    \label{fig:hysteresis}
\end{figure}

The BIPM Kibble balance group performed an experimental evaluation of the bias introduced by the yoke hysteresis \cite{hysteresis}. First, they  measured the magnetic flux density change $\Delta B_y$ at an approximate $BH$ working point as a function of the yoke $H$ field change, $\Delta H_y$. With an interpolation, they were able to estimate the $\Delta B_y$ value for the actual $\Delta H_y$ as presented in (\ref{eq:MNL}).  An interesting conclusion reached in \cite{hysteresis} was that the nonlinear bias is largely independent of whether the yoke is heat-treated or not. The reason is that the change in $H$ field in the yoke is inversely proportional to the yoke permeability $\mu_r$, as shown in (\ref{eq:MNL}). In the linear range, $\Delta B_y\approx \mu_0\mu_r\Delta H_y$, the permeability cancels out, therefore, $\Delta B_y$ has a similar value. For the BIPM case ($N=1057$, $I=13.3$\,mA, $\delta_a=13$\,mm), the estimated $B$ field change is about $2\times 10^{-8}$, regardless whether yoke was heat treated or not. Here, we use (\ref{eq:MNL}) and convert the bias into the NIST-4 system, it yields 1/25 of the BIPM effect, i.e. $8\times10^{-10}$. 
 Using this value, the hysteresis effect as a function of $p$ is plotted as the blue dots in Fig. \ref{fig:hysteresis}. As in the previous section,  the effect starts to become significant for magnet systems with smaller volumes ($p\le1/3$). Experimentally, the nonlinear magnetic error can be measured by weighing different test masses and a value of $\varepsilon_2(1)=(1.4\pm1.4)\times10^{-9}$ was found by experimental measurement with the NIST-4 system \cite{NIST}. Note, that this result contains contributions of both the $BH$ nonlinearity and the hysteresis of the yoke. Fig.\ref{fig:hysteresis} shows also the experimental value and its scaling as a function of $p$.

\section{Coil Thermal Effect}
\label{sec05}
A typical sequence for a Kibble balance measurement is alternating weighing and velocity measurements. A slow drift of the geometric factor can then be removed by processing the data ABA scheme~\cite{swanson_2010}. Such processing only removes temporal changes of $Bl$ that are uncorrelated with the measurement mode of the experiment, caused, for example, by slow changes in the room temperature. Such drifts do not add a bias to the result, {but} could add random fluctuation and hence lengthen the measurement time required to achieve the desired uncertainty for the result.

A temperature change that is coherent with the measurement mode is different. It can yield a systematic effect because the $(Bl)$ in the mode where the temperature is higher, typically the weighing mode, will be lower.

Unfortunately, it is difficult to know if a temperature difference between the weighing and velocity measurements exists. It would be easy to mount a temperature probe on the magnet and record its reading during the measurement. However, due to the large thermal capacity of the magnet, the temperature measured at the magnet surface is delayed and attenuated from the temperature of the magnetic material. Hence, one has to be careful to interpret these measurements. Two strategies can mitigate a systematic temperature change: (1) minimizing the ohmic power loss in the coil, and (2) maximizing the thermal capacity of the permanent magnet system to keep the temperature rise small for a given power dissipation. Both measures are automatically implemented in a conventional large-volume magnet system. As we shall see both measures become crucial when the magnet volume or mass is reduced.

In this section, we study the thermal effect related to coil heating during the weighing measurement with finite element analysis (FEA). The NIST-4 system ($p=1$) is used as an example. Heat can be conducted through the yoke, the gaseous material in the gap, and radiation. For the simulation, the magnet is immersed in low-density air. The density is set as $\rho=1\times10^{-8}\,$kg/m$^3$, a value that is comparable to low vacuum in Kibble balance experiments. To easily visualize the effect, the coil heating  power in the weighing phase is exaggerated to 10\,W.  Initially, the system is assumed to be at 25\,$^\circ$C. 

\begin{figure}
    \centering
    \includegraphics[width=0.5\textwidth]{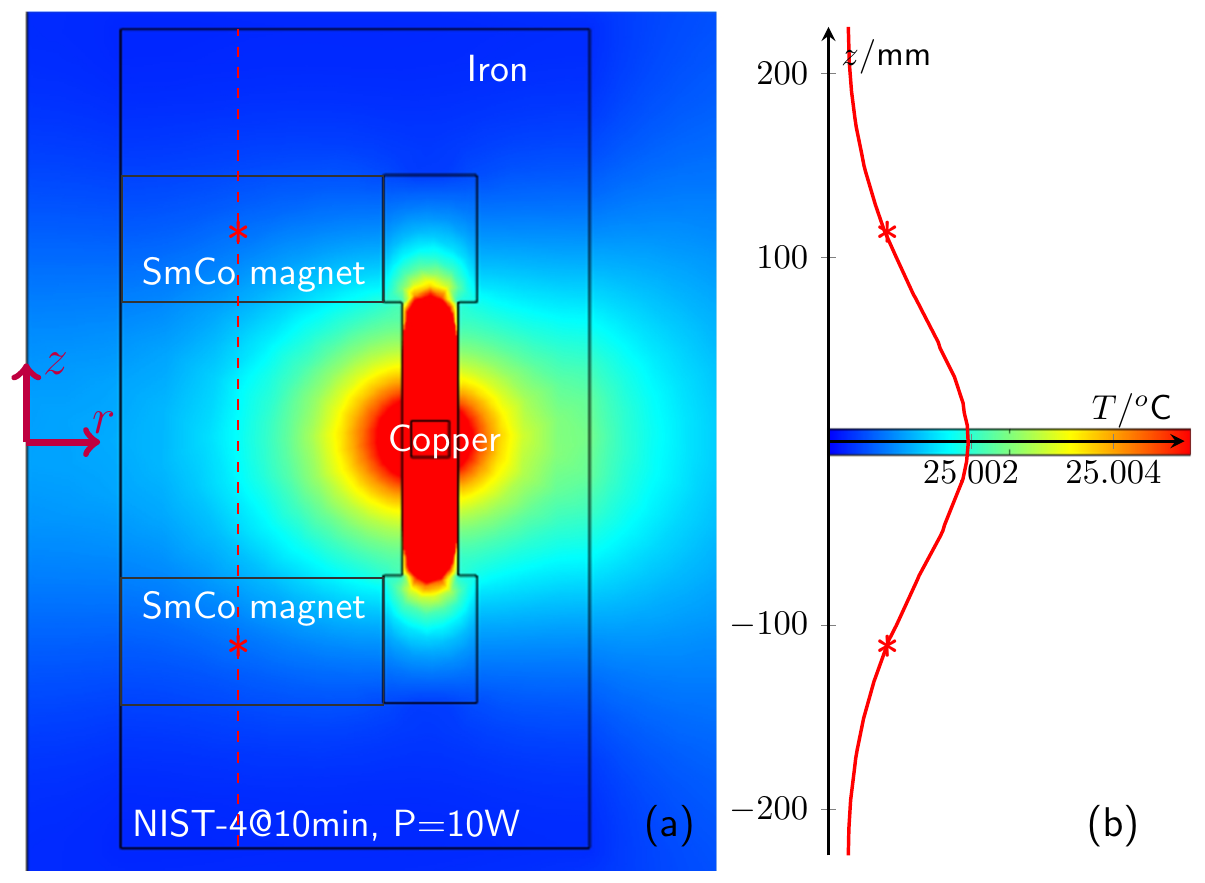}
    \caption{(a) Thermal distribution of the NIST-4 system. The plot shows the result after heating the coil for 10 minutes at a power of 10\,W. (b) The temperature distribution along a vertical dashed line in (a) is shown as the red line in (b). The color bar in (b) shows the scale of temperature map (a).}
    \label{fig:map}
\end{figure}

Fig. \ref{fig:map} (a) shows the thermal distribution in the magnet system after 10 minutes of heating through the coil. Although a hot spot with a maximum temperature change of 2.9\,K has developed at the coil,  the thermal change at the center of the magnetic material (SmCo,  marked by * in Fig. \ref{fig:map}) is only 0.89\,mK, due to the large thermal capacity of the massive structure. Then, the scale factor $p$ is applied to the FEA model, while the heating power is kept constant at $P=10$\,W. With $p=1/2$, a temperature change of 30.5\,mK is obtained, and with $p=1/3$, the thermal change at the magnet center is 196.6\,mK. At fixed power, both results scale approximately as 
\begin{equation}
    \frac{\Delta T_\mathrm{mag,P}(p)}{\Delta T_\mathrm{mag,P}(1)}\approx \frac{1}{p^5}.
    \label{eq:thermal}
\end{equation}
As mentioned above, this result assumes the same power dissipation. Since most coils are wound with copper, the only parameters are the cross-section of the wire and its length, i.e., the number of turns times the coil circumference. As we scale the magnet the cross-sectional area of the air gap scales by $p^{2}$. We assume that the coil uses the same fractional space of the gap, hence the cross-section of the coil scales by the same factor. The cross-sectional area of the coil is the product of the number of turns and the area of the wire cross-section $s$. Hence,
\begin{equation}
    \frac{N(p)s(p)}{N(1)s(1)}=p^2\rightarrow \frac{s(p)}{s(1)}=p^3.
\end{equation}
With the above equation, it is easy to calculate the scaling law for the power loss. It is
\begin{equation}
    \frac{P_c(p)}{P_c(1)}=\frac{I(p)^2R_c(p)}{I(1)^2R_c(1)}=\frac{l(p)}{l(1)}\frac{s(1)}{s(p)}=\frac{1}{p^3},
    \label{eq:power}
\end{equation}
where $R_c$ is the coil resistance. {For reference, the coil resistance in the NIST-4 Kibble balance is $112\,\Omega$.} Note that the length of the coil does not change, because the number of turns is inversely proportional to $p$ and the coil's  circumference is proportional to $p$. Hence, in the product, the $p$ cancels. {As one changes the cross-section of the coil, the interwinding capacity will change. The experimenters should be aware of that. A difference in capacity might negatively affect the dynamic response of the system and could produce or change a bias due to dielectric relaxation, i.e, the voltage across the coil changes slowly due to a change in the dielectric permittivity of the wire insulation on a long time scale.}

Combining (\ref{eq:thermal}) and (\ref{eq:power}) yields a $Bl$ change due to the coil ohmic heating given by
\begin{equation}
    \frac{\varepsilon_3(p)}{\varepsilon_3(1)}=\frac{\alpha_m(p)}{\alpha_m(1)}\frac{\Delta T_\mathrm{mag,P}(p)}{\Delta T_\mathrm{mag,P}(1)}\frac{P_c(p)}{P_c(1)}\approx \frac{1}{p^8}.
    \label{eq:e3}
\end{equation}
Here, $\alpha_m$ is the temperature coefficient of rare earth magnets. For Sm$_2$Co$_{17}$, $\alpha_m\approx-3\times10^{-4}$/K.    

\begin{figure}[tp!]
    \centering
    \includegraphics[width=0.425\textwidth]{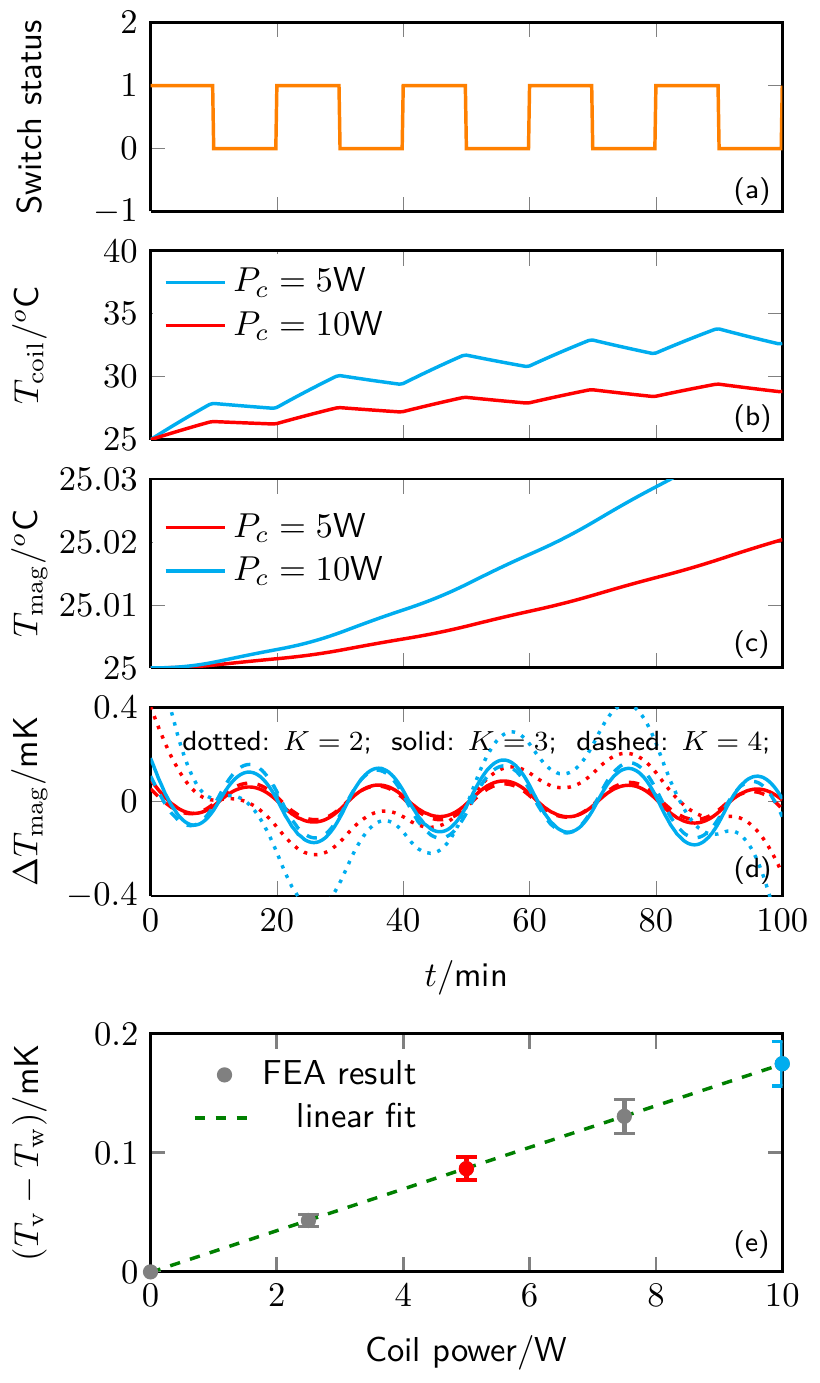}
    \caption{(a) presents the status of the coil power. A status of 1 means the heating is on as it is the case during weighing measurement. The velocity measurement contains no current and hence the heating status is 0. (b) and (c) show the temperature change over time respectively at the coil position and the magnet center with two different power levels in the coil, $P=10$\,W and $P=5$\,W. (c) shows the residual variations of the thermal change at the magnet position. The residual thermal change is defined as the FEA data minus its $K$-order polynomial fit, i.e. $\Delta T_\mathrm{mag}=T_\mathrm{mag}-\hat{T_\mathrm{mag}}(K)$. Two colors label the residual with different powers, similar to (b) and (c). (e) presents the average temperature difference in the weighing and velocity measurements as a function of the heating power. The error bar is obtained by the standard deviation of the last four periods of the weighing and velocity measurements. Note, all the results are obtained with the NIST-4 system with $p=1$.  }
    \label{fig:error3}
\end{figure}

Next, we estimate $\varepsilon_3(1)$  by introducing a modulated heating source in the FEA calculation. As shown in Fig.~\ref{fig:error3}\,(a), the heating source is periodically turned on (1) and off (0) with a period of 20 minutes, equally divided between heating-on (weighing measurement) and heating-off (velocity measurement). The temperature change at the coil position $\Delta T_\mathrm{coil}$ and the magnet position $\Delta T_\mathrm{mag}$ with heating power $P=10$\,W and $P=5$\,W are shown in Fig.\ref{fig:error3}\,(b) and (c), respectively. The results show both temperature changes are proportional to $P$. The coil temperature change $\Delta T_\mathrm{coil}$ shows a clear synchronization with the switch status, i.e. $\Delta T_\mathrm{coil}$ increases in each weighing measurement and decreases during velocity measurements. Due to the large thermal capacity, the increase of $\Delta T_\mathrm{mag}$ is much smoother. To obtain the thermal change, a polynomial fit is removed from the data. Fig.\ref{fig:error3}\,(d) shows the residual terms $\Delta T_\mathrm{mag}-\hat{\Delta T_\mathrm{mag}}(K)$ for three different orders $K$ of the polynomial drift subtraction. It can be seen that with $K=2$ the residual curve is not yet stabilized. There are only minor differences between $K=3$ and $K=4$, so, $K=3$ is sufficient to remove the slow temperature drift. An {interesting} result is a phase difference between the temperature at the permanent magnet position and the heating power. The temperature is delayed by 90 degrees. As a result, the average temperature during the velocity measurement $T_\mathrm{v}$ is greater than that during the weighing measurement $T_\mathrm{w}$.  Fig.\ref{fig:error3}(e) shows the  $T_\mathrm{v}-T_\mathrm{w}$ obtained with the same procedure explained above for five different power levels of the heater. The results are in excellent agreement with a linear relationship between temperature difference and dissipated power. The calculated slope is  $k_m=0.0175$\,mK/W.  The  power consumption of the  NIST-4 coil is $P_c(1)\approx 5.5$\,mW and we use a temperature coefficient of $\alpha_m=-3\times10^{-4}$/K. With that, all the necessary values are available to deduce $\varepsilon_3(1)$. It is
\begin{eqnarray}
    \varepsilon_3(1)&=&\frac{B_\mathrm{w}}{B_\mathrm{v}}-1=\alpha_m(T_\mathrm{w}-T_\mathrm{v})k_mP_c(1)\nonumber\\
    &\approx& 2.89\times10^{-11}. 
\end{eqnarray}
For $p=1$, the effect is truly negligible. Three orders below the uncertainty of the NIST-4 Kibble balance.

\begin{figure}
    \centering
    \includegraphics[width=0.5\textwidth]{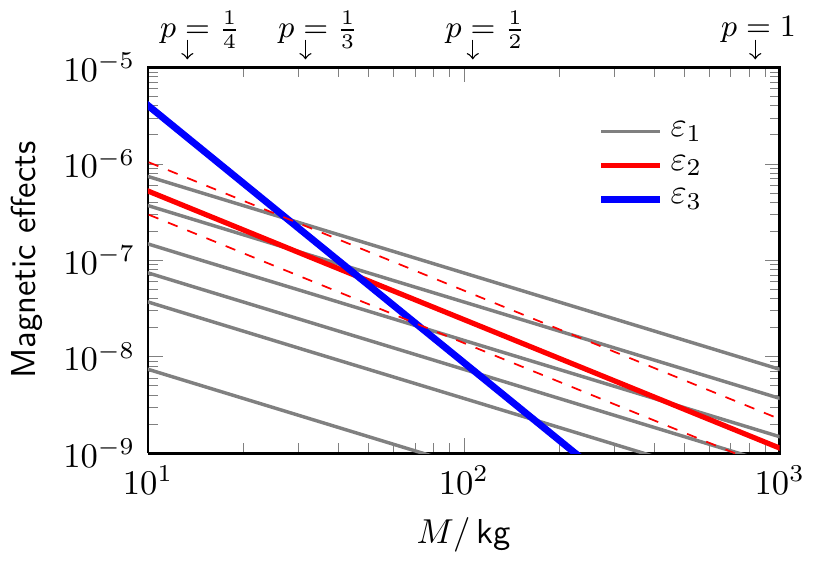}
    \caption{The coil ohmic heating effect as a function of $p$ in blue. The red and gray traces show the effects of the linear and nonlinear magnet errors have been plotted in Fig.~\ref{fig:e1} and Fig.~\ref{fig:hysteresis}. 
       }
    \label{fig:e3}
\end{figure}

While the effect is minuscule for $p=1$, the $p^{-8}$ dependence will lead to a strong increase with decreasing $p$.
The $\varepsilon_3(p)$ as a function of $p$ is plotted together with $\varepsilon_1(p)$ and $\varepsilon_2(p)$  in Fig.\ref{fig:e3}. The former effect raises sharply with decreasing magnet mass. For example, $\varepsilon_3({1}/{2})\approx7\times10^{-9}$ is below the Kibble balance uncertainty limit. However, for $p=1/3$ and  $p=1/4$, the effects are significant, $\varepsilon_3({1}/{3})\approx1.9\times10^{-7}$ and $\varepsilon_3({1}/{4})\approx1.9\times10^{-6}$.

\section{Scaling with mass of the test mass}
\label{sec06}
So far, we have assumed that the balance weighs a 1\,kg mass although the magnet system is scaled. Here we revisit this assumption since most tabletop Kibble balances aim at measuring smaller masses, typically in a range from 1\,g to 100\,g. We introduce a second scaling factor $q$. The mass that the Kibble balance measures is given by $m=q\cdot 1\,\mbox{kg}$. All equations discussed above imply $q=1$.

Eq. (\ref{eq:opBl}) shows that the optimal $Bl$ value is proportional to $\sqrt{m}$. Since $B$ is independent of $p$, the wire length $l$ can be reduced to  $\sqrt{q}~l$. Similar to arguments made above, the number of turns is then reduced to $N(p,q) =N(1,1)/(p\sqrt{q})$. 
The linear magnetic effect, as described by (\ref{eq:e1}) depends on the current and the number of turns. The former does not scale with mass, but the latter does. Hence, it is
\begin{equation}
\frac{\varepsilon_1(p,q)}{\varepsilon_1(1,1)}=\frac{\sqrt{q}}{p}
\end{equation}
Here, we assumed that $(z_\mathrm{on}-z_\mathrm{off})$ is independent of mass. In other words, the stiffness of the mechanical system is  proportional to the mass. 

The scaling of the nonlinear magnetic effect is given in ~{(\ref{eq:MNL})}. Since the effect is proportional to $\Delta H_y^2$, we find
\begin{equation}
\frac{\varepsilon_2(p,q)}{\varepsilon_2(1,1)}=\frac{q}{p^4}
\end{equation}

Finally, for the thermal effect, the reduction of coil turns leaves more space for wiring and hence the wire gauge increases, i.e. $s(p,q)=s(p,1)/\sqrt{q}$. The total resistive power loss is proportional to $l/s$ and, therefore, the heating power is reduced to $P_c(p,q)=q P_c(p,1)$. The same reduction is achieved for $\varepsilon_3$. It is,
\begin{equation}
\frac{\varepsilon_3(p,q)}{\varepsilon_3(1,1)}=\frac{q}{p^8}
\end{equation}

The magnetic errors ($\varepsilon_1,\varepsilon_2)$ can be kept constant by choosing $p^2=q$, i.e., the linear size of the magnet scales with the square root of the mass value of the test mass. If the mass is reduced by a factor of four the magnet can be made smaller by a factor of two.

Unfortunately, the thermal error becomes dominant for smaller $p$, and it requires $p^8=q$.  To keep $\varepsilon_3$ constant when scaling the magnet of a  balance that can weigh 1\,kg by a factor of four, the scaled system would only be able to weigh 15.3\,g. As a consequence, good thermal engineering is important when scaling the size of the magnet down.

\section{Mitigation of the scaling effects}
Shrinking the size of the magnet adversely affects the three uncertainties discussed in this article. A few measures can be taken  to improve this situation:
\begin{enumerate}
\item {\it Compensating the coil movement.} 
The inductance effect is proportional to the change in the coil position in the two weighing measurements, $z_\mathrm{on}-z_\mathrm{off}$. Hence, two solutions to reduce $\varepsilon_1$ are readily apparent. First, stiffen the suspension to reduce $z_\mathrm{on}-z_\mathrm{off}$. Second, measure the dependence of the result on $z_\mathrm{on}-z_\mathrm{off}$, by varying that position difference. This can be achieved in three ways, (a) changing the stiffness of the suspension, (b) weighing different masses, and (c) introducing a bias to the servo position of the balance beam. Independent of which of the three methods are used, the sensitivity of the result on $z_\mathrm{on}-z_\mathrm{off}$ can  be measured and a correction can be applied.

\item {\it Improving the magnet design.} 
The current effect is inversely proportional to the air gap width $\delta_a$. 
Enlarging the air gap width can reduce the inductance error and nonlinear error. Unfortunately, it also reduces the vertical region in which the magnetic field is uniform.
Therefore, a compromise must be found. A novel design idea to compensate for the edge effect allows to enlarge the uniform area and still widen the gap~ \cite{li2020simple}. Such and similar ideas can reduce the thermal effect.

\item {\it Employing new material.} 
The most direct way to reduce $\varepsilon_3$ is to find a material that has a lower temperature coefficient $\alpha_m$.  New rare-earth materials can be used, for example Sm$_2$Co$_{17}$Gd \cite{marangoni2019magnet}. This material has a temperature coefficient of $10^{-5}/$K, about 30 times smaller than conventional SmCo. Another ingenious solution that can reduce the temperature coefficient of the entire magnet system is the introduction of a flux shunt~\cite{METAS}. A flux shunt shorts some of the magnetic flux so it is not available for the gap. As the temperature increases, the shunt becomes less effective shorting the flux. If designed correctly, the flux available in the gap can be made largely independent of $T$. Both methods have a common drawback. Less flux is available for the gap and hence, the current in the coil has to be made larger to produce the same amount of force. The heating power scales with the current squared, but only linear with the temperature coefficient. So for this to work, the improvement in $\alpha$ has to be more than the reduction in $B^2$.

\item {\it Switching to the one-mode measurement scheme}
All three effects discussed in this article are related to currents in the coil. These exist only during the weighing measurement. In the one-mode measurement scheme, the current is present in the velocity phase, also. Hence, the $(Bl)_w$ and $(Bl)_v$ are both measured with the current present and are subject to the same changes.  The changes include both the thermal and demagnetizing effects of the currents. Therefore, the one-mode scheme can greatly reduce the bias that these effects can cause. 

One important point to recognize is that the $Bl$ gradient along the vertical obtained by the velocity measurement is twice as large as that obtained by the weighing measurement. For a detailed explanation and discussion, see\cite{li17}.
\end{enumerate}

\section{Conclusion}
We presented a first study of the influence of the magnet size on the relative systematic effects. 
The three effects that were investigated depend strongly on the normalized magnet size $p$:
current effect: $\varepsilon_1\propto p^{-3}$, hysteresis effect: $\varepsilon_2\propto p^{-4}$,  thermal effect: $\varepsilon_3\propto p^{-8}$. Since the thermal effect has such a strong dependence on $p$ it becomes the dominant effect for magnets below a total mass of 20\,kg. Hence, for magnets of this size and smaller, good thermal engineering is very important. An effective solution to all three effects is to employ the one-mode measurement scheme instead of the traditional measurement scheme that uses two separate modes for velocity and weighing measurement. The one-mode scheme, however, requires better alignment over the vertical range, and the current in the velocity phase introduces additional electrical measurement noise which enlarges the type A uncertainty. Ideally the two uncertainty components are balanced.

\section*{Acknowledgment}
The authors would like to thank Darine Haddad from NIST and Hongyu Sun from Tsinghua University for valuable discussions. 

\end{document}